\documentclass[conference]{IEEEtran}
\usepackage{multirow}
\usepackage{url}
\usepackage{cite}
\usepackage[cmex10]{amsmath}
\usepackage{array}
\usepackage{setspace}
\usepackage{dblfloatfix}
\usepackage{float}
\usepackage{xcolor}

\ifCLASSINFOpdf
   \usepackage[pdftex]{graphicx}
\else
\fi
\begin{document}
%
\title{Extreme Scaling of Supercomputing with Stranded Power: Costs and Capabilities}

\author{\IEEEauthorblockN{Fan Yang and Andrew A. Chien*}
\IEEEauthorblockA{Department of Computer Science, University of Chicago\\
\{fanyang, achien\}@cs.uchicago.edu\\
*also Math \& Computer Science, Argonne Natl Lab}
}

\maketitle

\begin{abstract}

Power consumption (supply, heat, cost) and associated carbon emissions
(environmental impact) are increasingly critical challenges in scaling
supercomputing to Exascale and beyond.  
We proposes to exploit stranded power, renewable energy that
has no value to the power grid, for scaling supercomputers,
Zero-Carbon Cloud (ZCCloud), and showing that stranded power can be
employed effectively to expand computing \cite{YangChien15}.  

We build on those results
with a new analysis of stranded power, characterizing temporal,
geographic, and interval properties.  We simulate production
supercomputing workloads and model datacenter total-cost-of-ownership
(TCO), assessing the costs and capabilities of stranded-power based
supercomputing.  Results show that the ZCCloud approach is
cost-effective today in regions with high cost power.  



The ZCCloud
approach reduces TCO by 21-45\%, and improves cost-effectiveness up
to 34\%.  We study many scenarios.  With higher power price, cheaper
computing hardware and higher system power density, benefits rise to
55\%, 97\% and 116\% respectively.  
Finally, we study future extreme-scale systems,
showing that beyond terascale, projected power requirements in excess of
100MW make ZCCloud up to 45\% lower cost, 
for a fixed budget, increase peak PFLOPS achievable by 80\%.







\end{abstract}


%
\IEEEpeerreviewmaketitle

\section{Introduction}


With the end of Dennard scaling \cite{BorkarChien11,exascale-challenges}, power
is an increasingly important concern for supercomputing
centers.  Today's 10-petaflop systems draw more than 10 MW
\cite{TianHe-2,BlueWaters-Power}, and near-exascale machines announced by
the US DOE will exceed 15MW \cite{Aurora,Summit}, with Exascale systems
now expected to exceed 25MW, despite earlier lower projections \cite{exascale-challenges}.
Numerous centers already operate under power ``caps'' due to
local utility contracts or concerns about carbon footprint \cite{Super-Demand-Response14}. 
In other recent cases, centers have had forced power reductions due to
regional and national emergencies \cite{Japan-Earthquake}.
Thus, there is broad agreement that power, including both 
its direct cost, as well as implied costs in cooling, facilities, and the associated environmental
impact \cite{BlueWaters-facility,Nersc-CRT-move,Gore07} are a major challenge for extreme-scaling of supercomputers.


Recent research explores new approaches to scaling
  supercomputing and datacenters
  \cite{Google-offshore,Microsoft-underwater}.  
  Our previous work proposes to power datacenters exclusively based
  on stranded power -- uneconomic or wasted power generated from
  variable renewable generators such as wind, called ZCCloud
  \cite{chien2015zero,SHPC,YangChien15}. The ZCCloud approach exploits
  power that in ``grid markets'' is not worth transmission and use.
  Such ``stranded power'' (SP) occurs in contemporary power grids due
  to renewable power generation variation and geographic dispersal.
  Such power is easily identified in market-based dispatch systems --
  it has a negative price -- the grid would literally pay the producer
  to not generate the power.

Stranded power is a remarkably large, untapped
  resource.  For example, in 2014 the MISO power grid curtailed 2.2
  terawatt-hours (TWh) of power and 5.5 TWh at negative price for a
  total of 7.7 TWh of stranded power from wind resources alone
  \cite{Bird2013,CurtailmentUS14,YangChien15}.  
  We analyzed 70 million transactions from a 28-month
  period for the Midcontinent Independent System Operator (MISO) power
  market, one of the largest power grids in the United States, serving
  42 million people and clearing \$37B of power annually.  The
  detailed characterization showed that stranded power has sufficient
  availability (up to 80\% duty factor) and quantity (100's of MW) to
  support large-scale computing, and demonstrated that stranded power
  can support a production supercomputing workload with good
  productivity.

We build on those results, 
combining new cost analysis (TCO) with prior productivity results to create
a comparison of cost-effectiveness.  First, we 
present
models for stranded power that characterize its temporal, geographic,
and interval properties.  Next, we combine those insights
with a TCO model from \cite{berral2014building} and study varied supercomputer
scaling scenarios, varying size and approach.
ZCCloud computing resources
are co-located with renewable generation and deployed incrementally
in single or groups of containers.  This approach eliminates building costs, enables free-cooling to lower
cooling costs, and easy scaling.  Co-location eliminates power transmission.
Specifically, we compare the productivity and cost of each system
type under a variety of power cost, compute hardware cost, and
system power density scenarios.
Overall, these studies
provide a perspective on the geographies and scenarios where ZCCloud
is attractive.
Finally we examine extreme scaling at Exascale, and
beyond.  Specific contributions include:




\begin{enumerate} 
  \item Compare supercomputer scaling
    expansion based on traditional and ZCCloud model (stranded
    green power as the only supply) showing
    these volatile resources can match the throughput of conventional
    approaches.

  \item Show that ZCCloud-based scaling produces nearly 50\% lower
    capital cost and 34\% greater cost-effectiveness (throughput/\$) in 
    a basic scaleup comparison

  \item Varying power price, we show that ZCCloud is already attractive
     in regions with high-cost power today, achieving 55\% better cost-performance.

  \item Exploring competing future hardware pricing trends: End of Moore (commoditization) and 
    Difficult Scaling (per-chip cost with scaling), we show ZCCloud's advantage
    increases with cheaper hardware, up to 97\% more cost-effective at 0.25x cost.

  \item Exploring system power density (end of Dennard), show
     ZCCloud has 116\% better cost-effectiveness if power
     densities increase 5-fold (projected within ten years)

  \item Exploring Exascale and beyond, we show 
    that growing power and physical infrastructure costs enable
    ZCCloud to both reduce capital cost by 45\% and increase peak PFLOPS by 
    80\% under a fixed budget.
\end{enumerate}

The paper organization is as follows: Section \ref{sec:background}
introduces the background on renewables, power markets, and datacenter costs.
Section \ref{sec:approach} summarizes the ZCCloud approach and
describes the origin of stranded power.  We assess the capabilities of
ZCCloud in Section \ref{sec:capabilities}, study 
cost in Section \ref{sec:cost}, and cost-performance
in Section \ref{sec:cost-performance}. 
We explore extreme scaling in Section \ref{sec:exascale} and 
related work in Section \ref{sec:discussion}.  Section
\ref{sec:conclusion} summarizes and proposes directions
for future work.

\section{Background}
\label{sec:background}

In this section, we first introduce the growth of renewable power 
and operation of energy markets.  Then we summarize key elements in 
optimizing datacenter costs.

\noindent \textbf{Renewable Power} 
Fossil fuels are the largest source of carbon dioxide emissions, and
are widely
believed to be driving climate change \cite{Gore07}.
Such concerns drive the rapid development of renewable
energy generation. Growth of solar and wind is the most
rapid, comprising 5.2\% of US
electricity generation in 2014 \cite{renewable2014} with wind supplying 80\% of that.  In many states
renewable power levels of more than 10\% 
have already been achieved. For example, California is at 20\% renewable portfolio
standard (RPS) in 2010 \cite{rps-california}.  More ambitious RPS
 goals have been adopted: 50\% by 2030 
in California \cite{California-50RPS}, 25-31\% by 2025 in Minnesota, 
and 50\% by 2030 in New York.  President 
Obama released the ``Clean Power Plan'' in August 2015 that establishes 
the national standards to limit carbon pollution from power plants.  
In 2015, the U.S. Department of Energy Wind Program announced a 35\% RPS, 404GW 
goal for wind by 2050 \cite{WindVision2015}.

\noindent \textbf{Power Grid and Energy Market}
Modern ISO’s dispatch generation and price power generation
in real time.  For example, the Mid-continent ISO (MISO) market \cite{miso}
uses
5-minute intervals: generators offer power to the grid, and it 
prices their offered power
with locational marginal price (LMP), which varies by 
site and depends on transmission structure and supply-demand balance.
An important goal for power purchase is ``merit order'',
where lower prices have higher priority.   Scheduling 
power is difficult because generation and demand must be matched
instantaneously and in general power is not stored.
Any sudden change in demand or supply must be addressed.
So, to ensure stability or when there is overgeneration or
transmission congestion, the result is stranded power.  Information
and Communication Technologies (ICT) are
the fastest growing use of power \cite{greenpeace15}.


\noindent \textbf{Datacenter Cost and Efficiency}
Today's peta-scale supercomputers are comprised of
thousands of servers, requiring 10's of megawatts of electricity, and thousands of sqft.
To accommodate, datacenters 
provide physical space, cooling, and power.
Building and operating datacenters is expensive, reaching
10 to 100 million dollars per year \cite{mira, Aurora, Summit}.  
With the goal of maximizing processing, managing costs requires a focus on cost 
efficiency \cite{hoelzle2009datacenter}.  Dominant
elements of cost are compute servers and datacenter facilities, 
comprising more
than 75\% of total cost of ownership (TCO) \cite{hoelzle2009datacenter}.  
Other cost factors include power, system administration, power transmission, and 
data networking. \cite{le2011intelligent,wang15grid,berral2014building}

\section{Supercomputing with Stranded Power}
\label{sec:approach}
\subsection{ZCCloud Approach}

To address growing
power, physical space,
and cooling requirements, supercomputer centers regularly build or expand
power infrastructure, machine rooms, and cooling
 \cite{YangChien15} at costs of millions of dollars.
We study Zero-Carbon Cloud (ZCCloud), 
a new approach that exploits stranded power
that is not accepted by the grid.  ZCCloud deploys computing resources in decentralized
containers on wind generation site, exploiting stranded power to
escape these limits and scale
supercomputers to higher levels of capability and performance, 

\begin{figure}[b!]
\centering
\vspace{-0.1in}
\includegraphics[width=2.5in]{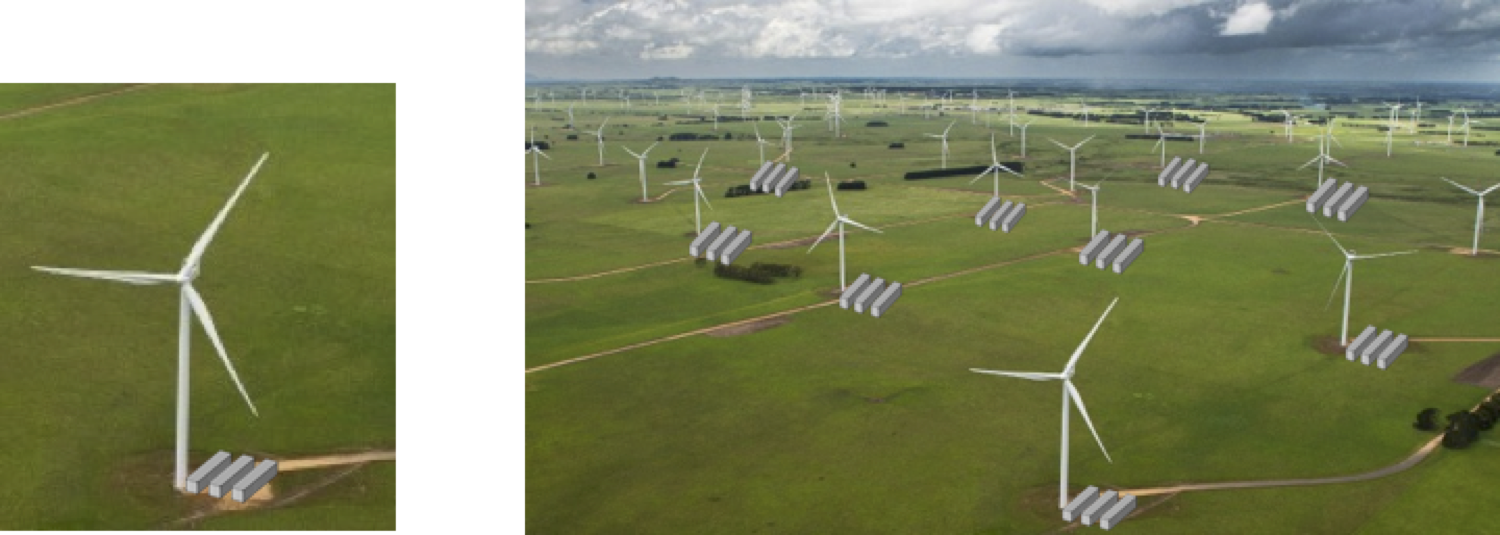}
\vspace{-0.1in}
\caption{ZCCloud containers below wind turbines.  Turbines can be several MW, and wind farms $>$250 MW.}
\vspace{-0.08in}
\label{fig:zccloud-on-farm}
\end{figure}

Power grids must balance match generation and load in real-time
However, intrinsic variability of renewable generation
(wind, solar, etc.)
creates major challenges.  To ensure
reliable power power, there is often oversupply and
transmission congestion that prevents generated power from reaching
loads. 


Stranded power is excess generation that cannot be
admitted to the power grid (curtailment) or is accepted
at negative prices (uneconomic power).
ZCCloud connects computing resources directly
to generators, avoiding congestion, and providing
direct access to stranded power.
This approach also minimizes
infrastructure for power transmission and distribution,
eliminates building costs, and enables free-cooling, and easy scaling.  
ZCCloud resources are only active when stranded power is available,
so they do not compete with normal grid loads.
Figure \ref{fig:zccloud-on-farm} shows a ZCCloud deployment on a wind
farm.  ZCCloud's modular approach
enables simple scaleup by deploying more containers.
Because modern commercial wind farms are 300 MW or larger, significant scale
computing resources can be deployed on a single site, simplifying
networking and administration.

\begin{figure}[t!]
\centering
\vspace{-0.1in}
\includegraphics[width=3in]{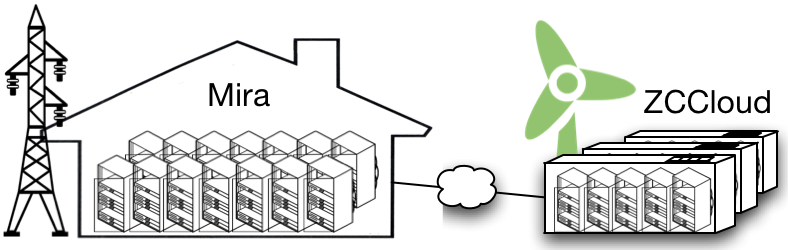}
\vspace{-0.1in}
\caption{System overview of Mira-ZCCloud.}
\vspace{-0.08in}
\label{fig:mira-zccloud}
\end{figure}

ZCCloud uses stranded power
as its only power source, and thus only operates when stranded power is
available and shut down when there is no stranded power.  
Short-term energy storage (e.g. batteries) and SSD's 
are used to enable checkpointing and workload migration within a short period 
before ZCCloud downtime.  We do not deploy sufficient energy storage that 
guarantees always-on service, because energy storage today is extremely 
expensive (more details will be discussed later).
Since ZCCloud resources are intermittent, we deploy them 
as a complementary extension, rather than a replacement of 
supercomputers such as ALCF’s Mira system (see Figure
\ref{fig:mira-zccloud}).  ZCCloud resources are paired with 
traditional datacenter supercomputing system such as Mira, using
a high-speed network to share a single filesystem, workload and scheduler.  
During ZCCloud uptime, jobs submitted to this mixed system are assigned to 
either traditional resources or ZCCloud resources.  When ZCCloud is off, 
jobs go only to the datacenter supercomputing resources.

\subsection{Stranded Power}
\label{sec:stranded_power}

Power grid scheduling attempts to match production to demand, but the
variability of power generation resulted from a growing cadre of renewable
sources (wind, solar) is a big challenge 
to effectively utilize and maintain grid stability \cite{Bird2013,E3report}.  When
overproduction occurs due to ramp or transmission congestion limits,
modern power markets will price the excess power with low or negative
locational marginal price (LMP).  In response, generators will avoid
putting power into the grid -- even discarding it -- to minimize loss.
Such reductions are called ``curtailment'', ``spillage'', or ``dispatch down''.  
We define \emph{Stranded Power} (SP) as all offered generation with no economic 
value, thus including both spillage and delivered power with zero or negative 
LMP.  

\begin{figure}[t!]
\centering
\vspace{-0.1in}
\includegraphics[width=2.5in]{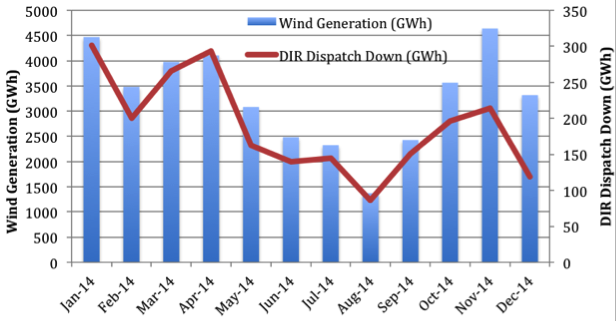}
\vspace{-0.1in}
\caption{Wind generation and down dispatch (MISO, 2014).}
\vspace{-0.08in}
\label{fig:miso_dispatch}
\end{figure}

Nationally, wind generation is 7x solar, so we focus on wind power.  
Figure \ref{fig:miso_dispatch} shows the monthly wind generation and
dispatch down of Mid-continent Independent System Operator (MISO).
In 2014, the total dispatch down was 2.2
terawatt-hours, and 183 MW in average, 7\% of wind generation.
Comparable waste exists in other regions of the US
\cite{rps-california}, and many European countries such as Denmark,
Germany, Ireland, and Italy \cite{Lew2013}. 
Waste is projected to increase
with higher renewable portfolio standard (RPS) 
\cite{E3report,California-50RPS}.

We quantitatively characterize stranded power by analyzing
MISO's real-time offers \cite{miso} (
LMP and generation data in 5-minute increments from 1/1/2013 to
4/14/2015).  The MISO market manages power for 1,259 generations site,
200 of which are wind sites, across more than 10 states in the US.
Wind power is the largest source of renewable power in MISO and accounts for
10\% of production.

We focus on two properties of stranded power: quantity and temporal
distribution.  Quantity is reported in megawatts or megawatt-hours, 
and temporal distribution as 
\textit{duty factor} -- the fraction of time when stranded power is
available.  Two families of stranded power models are used:

\begin{enumerate}
 \item \underline{Instantaneous Stranded Power} (\textit{LMP\{C\}}): 
 \textit{LMP$<$C} \\ ~~~~$(C=\$0, \$1, ..., \$5)$ for a 5-minute interval.
 \item \underline{NetPrice Stranded Power} (\textit{NetPrice\{C\}}): 
 \textit{NetPrice$<$C} \\ ~~~~ $(C=\$0, \$1, ..., \$5)$ for a period. 
\end{enumerate}
\begin{equation}
NetPrice = \frac{\sum_{period}LMP\cdot Power}{\sum_{period}Power} 
\end{equation}

We pick the sites with highest duty factors for each SP model, and
characterize their quantities and duty factors of stranded power.

\begin{figure}[t!]
\centering
\vspace{-0.1in}
\includegraphics[width=2.5in]{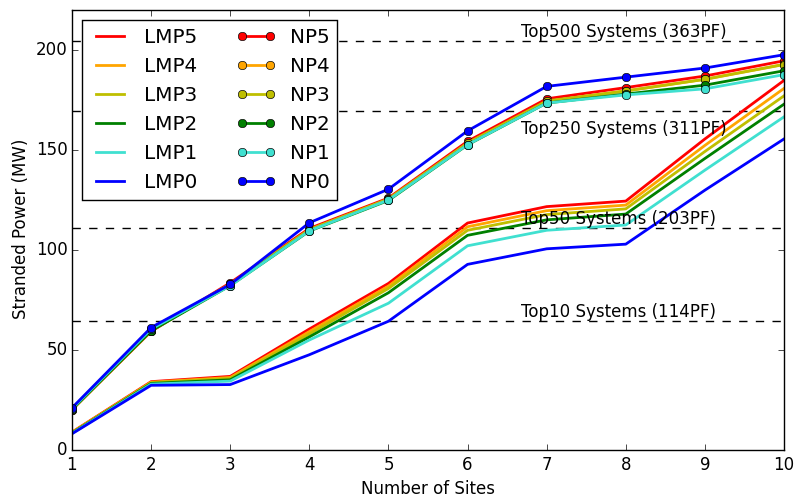}
\vspace{-0.1in}
\caption{Average Stranded Power vs. Number of Generation sites with highest duty factors, compared Top500 supercomputing systems (NP=NetPrice).}
\vspace{-0.08in}
\label{fig:stranded_power}
\end{figure}

\noindent \textbf{Quantity} Our studies show the quantity of
stranded power can be large.  Figure \ref{fig:stranded_power} plots
stranded wind power under different definitions with various numbers
of sites.  It compares the stranded power available (MW) to the total
power consumption of Top500 supercomputing systems \cite{top500}.
With the LMP model, two sites can support the Top 2 systems and five
sites are enough for the Top 10 systems.  For NetPrice (NP), stranded power
from a single site can support the top supercomputer (20MW); only
two sites are needed for the top 10 systems.  Of course, there is 
more stranded power in other regions beyond MISO.

\begin{figure}[b!]
\centering
\vspace{-0.1in}
\includegraphics[width=3.3in]{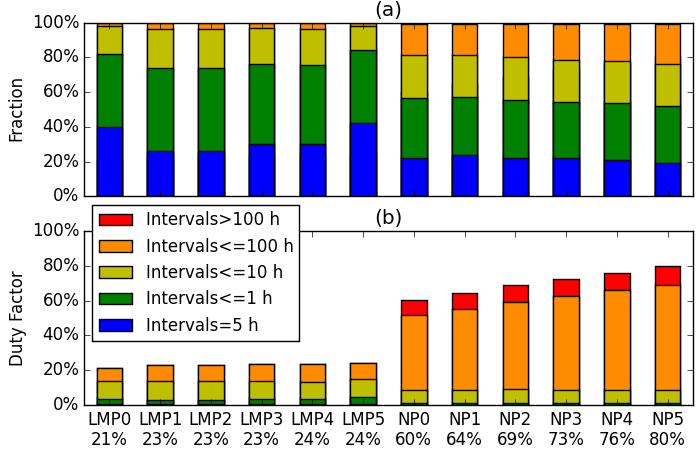}
\vspace{-0.1in}
\caption{
(a) Fractions of SP intervals by size for the best site.
(b) The contributions of each size SP interval to duty 
factor.  Results for a range of stranded power models. The percentage below 
each is the duty factor.}
\vspace{-0.08in}
\label{fig:intervals_top1}
\end{figure}

\noindent \textbf{Temporal Distribution}
We characterize the time periods when stranded power is available,
calling them SP intervals.  Figure \ref{fig:intervals_top1} shows 
the distribution of SP intervals for different SP models.
For LMP, 70\% of SP intervals are shorter than 1
hour, and overall
duty factors are less than 30\%.  For NetPrice the situation
is much better with half of the SP intervals 
$>$1 hour.  These longer intervals provide 
an overall duty factor between 
60\% and 80\%.  The longer intervals and greater duty 
factor arise from NetPrice's masking of brief fluctuations in LMP.

\begin{figure}[t!]
\centering
\vspace{-0.1in}
\includegraphics[width=3.4in]{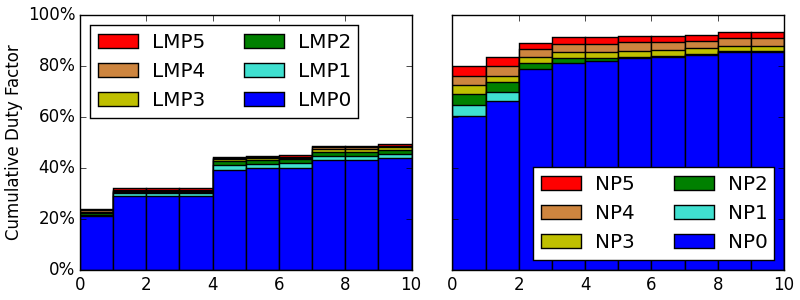}
\vspace{-0.1in}
\caption{Cumulative Duty Factor vs. Number of Generation sites with highest duty factor.}
\vspace{-0.08in}
\label{fig:cumulative_duty_factor}
\end{figure}

\noindent \textbf{Combined Temporal Distribution}
Is it possible to further improve duty factor by combining multiple
sites?  We calculate the cumulative duty factors across sites.
(see Figure \ref{fig:cumulative_duty_factor}).  For LMP,
more generation sites increases cumulative duty factor slowly,
reaching 30\% for 2 sites, and 50\% for 7 sites.  NetPrice yields much
higher duty factors of more than 80\% for only three sites.  However,
since there are periods with zero stranded power in the grid, it  
appears impossible to achieve a 100\% duty factor even with a large
number of sites.  

\noindent \textbf{Is 100\% Duty Factor Achievable with Storage?}
Likewise, employing sufficient energy storage to achieve 100\% duty factor is uneconomic.
MISO studies show periods without stranded power can be as long as 300 
hours, requiring more than \$400M energy storage for 4MW, approximately 4x the computing hardware cost  \cite{TeslaPowerWall}.
Coverage for 24-hour period is more than \$40M, 
nearly 1/2 the computing hardware cost.


In conclusion, stranded power from one or a few wind generation sites is
sufficient to power large supercomputers.  Such power can be available for 
up to 80\% of the time.  The results are encouraging for
stranded power-based computing.

\section{Capabilities of Intermittent Resources}
\label{sec:capabilities}

To assess the performance benefit of adding intermittent computing
resources to traditional datacenter, we simulate HPC workloads
first with a daily periodic model, and then with a more complex
set of uptime intervals derived from a single wind generation site's 
stranded power -- derived from real
MISO power grid statistics.  We vary system configurations, and compare
system throughput.

\subsection{Methodology}

\noindent \textbf{System Model} We use ALCF's Mira system \cite{mira}
as a model for traditional datacenters resources, a
10 PFlop IBM BG/Q system with 49,152 nodes, 786,432 cores, and 768 TB
memory.  Intermittent resources (ZCCloud) are connected to datacenter
(\textit{Ctr}) as an extension, denoted “\textit{Ctr+Z}”.  We compare
Ctr+Z with datacenter-only system (\textit{Ctr}).  Moreover, we
consider ZCCloud extension at varied scale: 1x, 2x, and 4x
of Mira resources, denoted \textit{Ctr+1Z, Ctr+2Z, and Ctr+4Z}.
For fair comparison, we also scale datacenter resources to 2x, 3x,
and 5x of Mira, denoted \textit{2Ctr, 3Ctr, and 5Ctr}, and use
\textit{1Ctr} as baseline.  Note that Ctr+\{n\}Z is of the same
size as \{n+1\}Ctr.

\noindent \textbf{Job Scheduling and Workload} We simulate ALCF job trace \cite{mira} 
using Mira's job scheduler, Cobalt v0.99 integrated with a simulator, Qsim 
\cite{cobalt}.  Workload properties are summarized in Table \ref{table:alcf_trace}.  
Workloads are scaled to match utilization on Mira, 
adding jobs with the same distributions of attributes (job size is not increased).  
On Ctr+Z configuration, the scheduler assigns 
jobs equally to datacenter and ZCCloud resources when ZCCloud is 
available, and only to the datacenter when ZCCloud is shutdown.
Because the NetPrice models produce stranded-power dominated by
intervals of 10 hours or more (longer than most runs), our
simulations give the 
the job scheduler information on the length of ZCCloud intervals,
scheduling only jobs that can complete before ZCCloud 
shutdown. 

\begin{table}[tb]
\renewcommand{\arraystretch}{1.3}
\vspace{-0.1in}
\caption{ALCF Workload Trace}
\label{table:alcf_trace}
\vspace{-0.1in}
\centering
{\small
\begin{tabular}{l|p{5.6cm}}
\hline
Parameters & Values\\
\hline \hline
\# Jobs & 78,795\\
\hline
Time Period & 12/31/2013\textemdash 12/30/2014\\
\hline
Runtime (hrs) & 0.004\textemdash 82, Avg 1.7, StDev 3.0\\
\hline
\# Nodes & 1\textemdash 49,152, Avg 1,975, StDev 4,100\\
\hline
Rsc Utilization & 84\% of Mira (100\% availability)\\
\hline
App Domains & Physics, Chemistry, Material Science, CompScience, Biology, Engineering\\
\hline
\end{tabular}
}
\vspace{-0.08in}
\end{table}

\noindent \textbf{Intermittent Resource Models}
Two resource models: (1) Periodic
intermittent resources: power is available in a fixed daily cycle,
(e.g. 8:00 to 20:00 each day), and  (2) Stranded-Power-based (SP-based)
resources: single-site stranded power history from MISO
\cite{miso} as described in Section
\ref{sec:stranded_power}. To normalize between them, we 
use
\textit{duty factor}, the 
fraction of time resources are available.
For example, a
periodic model with uptime 8:00-20:00 has 50\% duty factor, and
an SP-based model using LMP0 has 21\% duty factor.

\noindent \textbf{Metrics} 
Throughput (number of jobs per day) quantifies system capability.

\subsection{Periodic Intermittent Resources}
\begin{figure}[b!]
\centering
\vspace{-0.1in}
\includegraphics[width=2.5in]{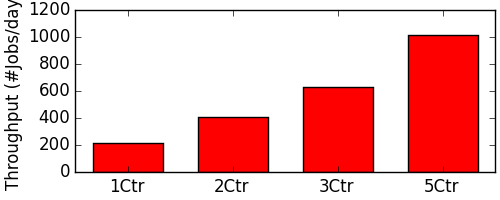}
\vspace{-0.1in}
\caption{Throughput vs. System size for Ctr.}
\vspace{-0.08in}
\label{fig:throughput_ctr}
\end{figure}

We first explore the scaling of traditional datacenter resources (see 
Figure \ref{fig:throughput_ctr}).  The system throughput of
Ctr scales linearly with resources, reaching
throughput of 200 jobs/day (1Ctr), and 1000 jobs/day with 5x resources
(5Ctr).

\begin{figure}[tb]
\centering
\vspace{-0.1in}
\includegraphics[width=3in]{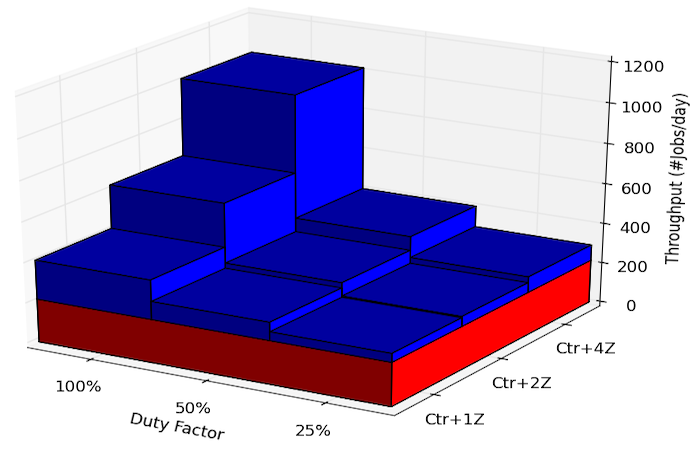}
\vspace{-0.1in}
\caption{\underline{Periodic Resources:} Throughput vs. Duty factor vs. System size for Ctr+Z (Blue) and Throughput of 1Ctr (Red).}
\vspace{-0.08in}
\label{fig:throughput_periodic}
\end{figure}

Next, we evaluate the benefits of periodic intermittent resources, varying duty factor and 
system size (Figure \ref{fig:throughput_periodic}).
System throughput of Ctr+\{n\}Z is greater than that of 
1Ctr in all cases, but less than 
\{n+1\}Ctr because of its frequent downtime.  As expected, at a duty factor of 100\%, in 
which ZCCloud resources are always available, the capability of Ctr+\{n\}Z 
matches \{n+1\}Ctr.

For periodic resources, increasing duty factor improves throughput more 
significantly than adding intermittent resources.  
For example, while \{Ctr+1Z, 50\% duty factor\} and \{Ctr+2Z, 
25\% duty factor\} both provide 1.5x of 1Ctr node hours, the former achieves 
15\% higher throughput than the latter. 

\subsection{Stranded Power-Based Resources}
We consider four cases of SP-based resources, using different SP
models including LMP0, LMP5, NetPrice0, and NetPrice5.  As we have
described in Section \ref{sec:stranded_power}, the corresponding duty
factors are 21\%, 24\%, 60\%, and 80\% respectively.

\begin{figure}[tb]
\centering
\vspace{-0.1in}
\includegraphics[width=3in]{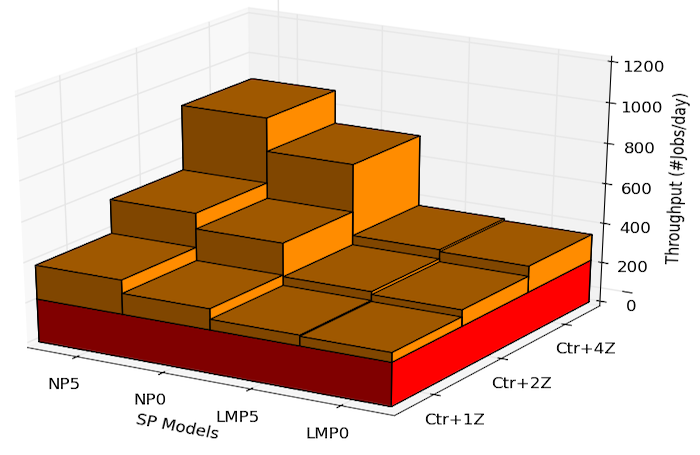}
\vspace{-0.1in}
\caption{\underline{SP-based Resources:} Throughput vs. SP model vs. System size for Ctr+Z (Orange) and Throughput of 1Ctr. (Red)}
\vspace{-0.08in}
\label{fig:throughput_zcc}
\end{figure}

Figure \ref{fig:throughput_zcc} shows the throughput of SP-based
resources for different SP models and scales.  Again, Ctr+Z achieves
better throughput than 1Ctr even for the worst case (LMP0, Ctr+1Z).
The much higher duty factor of NetPrice models ($>$60\%) creates big
advantage over LMP models.  For example, at the scale of Ctr+1Z,
NetPrice5 provides 1.8x of 1Ctr node hours while LMP5 only provides
1.24x.  
Throughput scales linearly 
with larger size and higher duty factor.  This is because most jobs are relatively 
small compared to the system scale and ZCCloud uptime.  For example, 
while job runtime is 1.7 hours in average and no longer than 82 hours, 
NetPrice5-based resources can provide uptime much longer than 100 hours. 
However, even with very high duty factor, Ctr+\{n\}Z is still less capable than
\{n+1\}Ctr with the same amount of computing resources.

Overall, adding intermittent computing resources to a datacenter
improves throughput, and the benefit is determined by duty factor
and scale, so intermittent resources of a given scale provide less capability
than traditional datacenter resources.

\section{Cost of Intermittent Resources}
\label{sec:cost}

We have shown supercomputing capability can be scaled up using
intermittent resources.  ZCCloud uses stranded power and containers,
lowering capital costs.  We describe a total-cost-of-ownership (TCO)
model for both datacenter and ZCCloud resources, applying it to quantify 
costs for different scaleup approaches.

\subsection{A Model for Total Cost of Ownership (TCO)}

We define a simple TCO model, starting from excellent academic and
industrial models \cite{berral2014building,wang15grid, le2011intelligent,hoelzle2009datacenter}, simplifying
by only 
including features that contribute more than 1\% to TCO.  For detailed
discussion see Appendix \ref{sec:tco-details}.

Our model includes several elements of cost: (1) computing
hardware (e.g. servers, racks, networking), 
(2) datacenter (DC) physical facilities, e.g. building,
raised floor, power distribution, etc., 
(3) electrical power,and (4) networking, the cost of fiber to an Internet backbone.
Thus, 
\begin{equation}
TCO(n) = n \cdot (C_{compute} + (C_{DCF} + C_{power})\cdot Density)) + C_{net} 
\end{equation}

Where these costs are parameterized as annual operating cost per unit 
of system, and $n$ is the number of units.  
We use Mira as the unit of supercomputing system, which is 4MW, 10PFlops
with a nominal cost of \$100M.
For example, if $n=2$, such as 2Ctr, the system is 8MW in total.  
$Density$ is a scale factor, used to model larger systems (in MW).


ZCCloud's TCO differs from traditional datacenters in three ways:  (5)
additional energy and SSD storage to enable checkpointing, 
(6) reduced DCF cost, by using containers and colocating at generation sites and
avoiding power distribution costs.  This reduces physical facility costs significantly.
(7) additional hardware for free cooling, 
and (8) ZCCloud uses stranded power, which is power that has no economic value in
the power grid.  We assume that this power can be used {\it at zero cost}, as this
is often a better deal than the negative prices that wind generators suffer.
Thus, the adjusted ZCCloud is follows:
\begin{equation}
TCO_z(n) = n \cdot (C_{z,compute} + (C_{ctnr} + C_{cool})\cdot Density) + C_{net} 
\end{equation}

\begin{equation}
C_{z,compute} = C_{compute} + C_{SSD} + C_{battery}
\end{equation}

where $C_{ctnr}$ is the cost of container per unit, and $C_{cool}$ denotes 
the free cooling hardware cost, amortized.



\begin{table}[t!]
\renewcommand{\arraystretch}{1.3}
\vspace{-0.1in}
\caption{Baseline Values in the Cost Model}
\label{table:tco_parameters}
\vspace{-0.1in}
\centering
{\small
\begin{tabular}{p{1.4cm}|p{1.8cm}|p{4.4cm}}
\hline
Parameters & Values & Source \\
\hline \hline
$C_{compute}$ & \$21M/unit & Mira \cite{mira} \\
\hline
$C_{DCF}$ & \$21M/unit & We assume $C_{DCF}$=$C_{compute}$ based on 
the case study in \cite{hoelzle2009datacenter} \\
\hline
$C_{power}$ & \$2.1M/unit & Annual electricity bill with US power price, 
\$60/MWh \cite{EIA-electricity} \\
\hline
$C_{net}$ & \$0.8M & From \cite{berral2014building} \\ 
\hline
$C_{SSD}$ & \$0.3M/unit & Intel SSD \cite{IntelSSD} \\
\hline
$C_{battery}$ & \$0.1M/unit & Tesla PowerWall \cite{TeslaPowerWall} \\
\hline
$C_{ctnr}$ & \$2M/unit & Vendor quote in \cite{chien2015zero} \\
\hline
$C_{cool}$ & \$0.3M/unit & Free cooling HW \cite{berral2014building} \\ 
\hline
$Density$ & 4MW/\$100M & Mira \cite{mira} is the Unit, (1x)\\
\hline
\end{tabular}
}
\vspace{-0.08in}
\end{table}



\begin{figure}[b]
\centering
\vspace{-0.1in}
\includegraphics[width=3in]{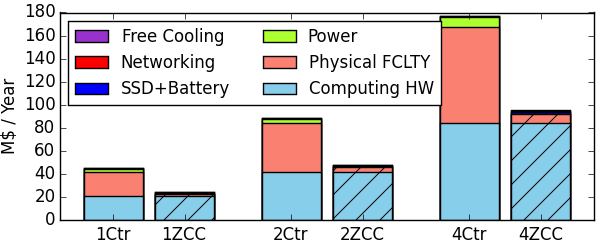}
\vspace{-0.1in}
\caption{TCO breakdown of Ctr and ZCCloud at same scale (1x, 2x, 4x).}
\vspace{-0.08in}
\label{fig:tco_breakdown}
\end{figure}

\begin{figure*}[t!]
\centering
\vspace{-0.1in}
\includegraphics[width=6.4in]{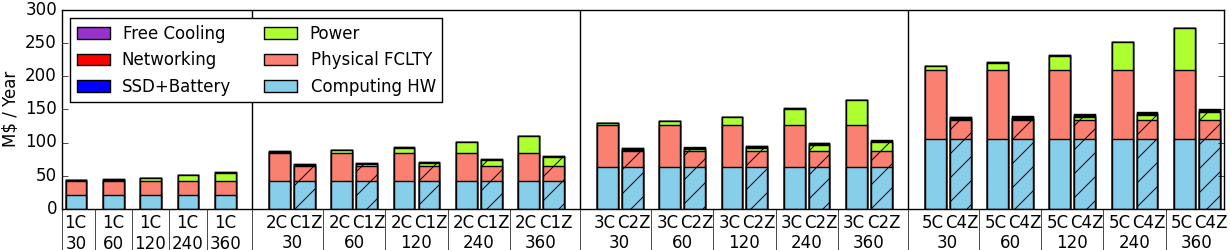}
\vspace{-0.1in}
\caption{TCO vs. System scale vs. Power price.  Bottom labels are power prices in \$/MWh.  
The system scale increases left to right.  We compare \{n+1\}Ctr and Ctr+\{n\}Z (\{n+1\}C = \{n+1\}Ctr, and C\{n\}Z = Ctr+\{n\}Z).}
\vspace{-0.08in}
\label{fig:tco_breakdown_power}
\end{figure*}

\begin{figure*}[t!]
\centering
\vspace{-0.1in}
\includegraphics[width=6.4in]{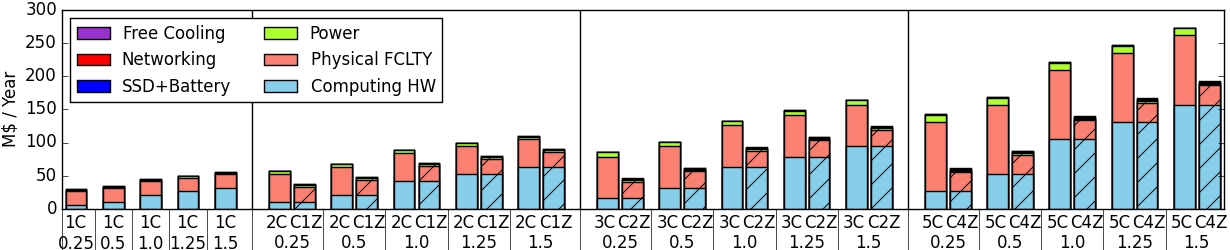}
\vspace{-0.1in}
\caption{TCO vs. System scale vs. Compute HW price.  
Bottom labels are scale factors of compute HW price.  
The system scale increases left to right.  We compare \{n+1\}Ctr and Ctr+\{n\}Z (\{n+1\}C = \{n+1\}Ctr, and C\{n\}Z = Ctr+\{n\}Z).}
\vspace{-0.08in}
\label{fig:tco_breakdown_compute}
\end{figure*}

From public documents and discussions with the Argonne
team \cite{ALCF}, vendor quotes for containers, and widely
accepted guidelines  \cite{hoelzle2009datacenter},
we derive the baseline values for the TCO 
model shown in Table \ref{table:tco_parameters}.
See 
Appendix \ref{sec:tco-details} for the full detail.

\subsection{Exploring TCO for ZCCloud Systems}
We calculate the costs of Ctr and ZCCloud systems for a range of
different sizes and under different TCO assumptions (see Table
\ref{table:study_space}).  In Figure \ref{fig:tco_breakdown}, we
illustrate our TCO model for the baseline costs (power price=\$60/MWh,
computing price=1x).  Savings in physical infrastructure by avoiding
power distribution and physical buildings, reduce ZCCloud resources
capital costs by nearly half.\footnote{Note that networking, SSD,
  energy storage, and free cooling costs are relatively small compared
  to other three major components.}  Power cost is eliminated, so
ZCCloud TCO is dominated by compute HW.

\noindent \textbf{TCO vs System Size and Power Price}
We illustrate the TCO model, scaling system size from 1-5x, and power price from \$30 
to \$360 / MWh  ($C_{power}$ ranging from \$1.1M/unit to \$12.6M/unit), matching
the current wide range of power prices across 
the globe.
We compare scaling traditional datacenters to ZCCloud,
showing resulting TCO's in Figure
\ref{fig:tco_breakdown_power}.  Within each group, the higher power prices
increase power cost of each system.   But power cost is not dominant.
The TCO of the Ctr+Z systems is less than
corresponding Ctr systems because of their lower 
physical facility costs.  At lowest power price, 
(\$30/MWh), Ctr+1Z is 21\% cheaper than 2Ctr, and at 
\$360/MWh, Ctr+4Z is 45\% cheaper than 5Ctr.  

\begin{table}[t!]
\renewcommand{\arraystretch}{1.3}
\vspace{-0.1in}
\caption{Cost-Performance Study Space}
\label{table:study_space}
\vspace{-0.1in}
\centering
{\small
\begin{tabular}{p{1.4cm}|p{6.4cm}}
\hline
Parameters & Values\\
\hline \hline
$\#Units$ &  n = 1, 2, 4 for \{n+1\}Ctr and Ctr+\{n\}Z \\
\hline
$C_{compute}$ & 5.2, 10.5, 21, 26.2, 31.4 M\$/unit (for compute price 
= 0.25x, 0.5x, 1x, 1.25x, and 1.5x)\\
\hline
$C_{power}$ & 1.1, 2.1, 4.2, 8.4, 12.6 M\$/unit (for power price 
= 30, 60, 120, 240, 360 \$/MWh) \\
\hline
$Density$ & 1x, 2x, 3x, 4x, and 5x\\
\hline
\end{tabular}
}
\vspace{-0.08in}
\end{table}

\begin{figure*}[t!]
\centering
\vspace{-0.1in}
\includegraphics[width=6.4in]{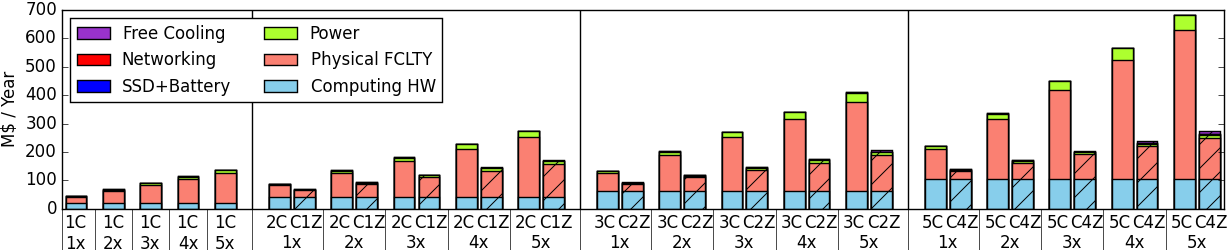}
\vspace{-0.1in}
\caption{TCO vs. System scale vs. Power Density.  
Bottom labels are system power densities.  
The system scale increases left to right.  We compare \{n+1\}Ctr and Ctr+\{n\}Z (\{n+1\}C = \{n+1\}Ctr, and C\{n\}Z = Ctr+\{n\}Z).}\vspace{-0.08in}
\label{fig:tco_breakdown_density}
\end{figure*}

\noindent \textbf{TCO vs System Size and Compute HW Price}
We explore the TCO model, varying the compute hardware cost (see
Figure \ref{fig:tco_breakdown_compute}) from 0.25x to 1.5x, a
range that encompasses commoditization (Intel processors at ARM prices)
or slight increase due to expensive technology scaling.
Scenarios with cheaper hardware increase the advantages of
ZCCloud 
because compute hardware depreciation dominates ZCCloud costs.
With 
computing hardware prices 0.25x of current level, Ctr+1Z is 34\% cheaper 
than 2Ctr, and
Ctr+4Z is 57\% cheaper than 5Ctr.
At high computing prices, ZCCloud benefits are reduced.
For example, with 1.5x computing price, the 
cost of Ctr+1Z is 18\% lower than 2Ctr, and Ctr+4Z is 
30\% lower than 5Ctr.

\noindent \textbf{TCO vs System Size and System Power Density}
Exascale studies project that future supercomputers
purchased for a given \$\$ budget will increase in power consumption
to the end of Dennard Scaling.  Their compute power increases
at a higher rate.
We consider the effect of this shifting balance on TCO 
(see Figure 
\ref{fig:tco_breakdown_density}), seeing that 
physical facilities and power cost grow significantly, amplifying the benefits of
ZCCloud.   With 1x density, Ctr+4Z is 37\% cheaper than 5Ctr, and it is 
60\% cheaper than 5Ctr for 5x density.

While the addition of ZCCloud resources reduces the total cost, the frequent 
downtime of ZCCloud also lowers the system capability.  Therefore, to 
systematically assess the benefit of adding ZCCloud to datacenter, we further 
study the cost-effectiveness of scaling with ZCCloud resources.

\section{Exploring Cost-Performance for ZCCloud}
\label{sec:cost-performance}

In this section, we explore the cost-performance of periodic resources
and SP-based resources, and then explore regional difference in power
price, different hardware prices as well as growing system power
density.  To evaluate cost-performance, we combine our resource
productivity experiments with the TCO cost models.  Throughout, we use
the metric {\it Throughput / Million \$}, which is work per unit
operating cost.

We first consider a baseline scenario, using a typical power price for the
United States, \$60/MWh ($C_{power}=\$2.1M/unit$), and Mira's compute hardware 
price ($C_{compute}=\$21M/unit$).  First we
consider a simple periodic intermittent resource, and then stranded-power-based systems with more
complex intermittency.

\begin{figure}[bt]
\centering
\vspace{-0.1in}
\includegraphics[width=3in]{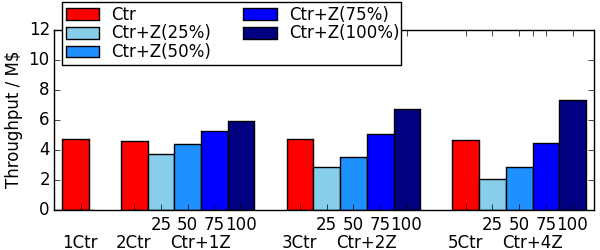}
\vspace{-0.1in}
\caption{Throughput/M\$ for a variety of System sizes and Duty factors.}
\vspace{-0.1in}
\label{fig:scaling_periodic}
\end{figure}

\noindent \textbf{Periodic Intermittent Resources}
We first explore periodic resources, varying duty factor from 25\% to 100\%.
The results show that as duty factor increases, cost-performance 
improves (see Figure \ref{fig:scaling_periodic}).
At 50\% or less, the ZCCloud systems achieve worse cost-performance than the 
traditional datacenter systems, 
because periodic intermittent resources cannot hold long-running jobs, 
shifting the heavy load onto traditional resources and making them overloaded.  
These
effects are magnified on the larger systems which have a larger proportion of ZCCloud
resources.  
As duty factor increases further, the ZCCloud systems becomes competitive 
with traditional datacenter systems.



\begin{figure}[bt]
\centering
\vspace{-0.1in}
\includegraphics[width=3in]{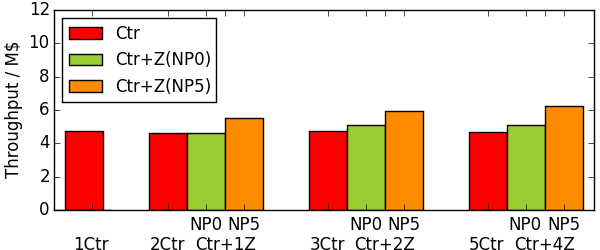}
\vspace{-0.1in}
\caption{Throughput/M\$ for a variety of System sizes and SP models.}
\vspace{-0.1in}
\label{fig:scaling}
\end{figure}

\noindent \textbf{Stranded-Power-Based Resources}
For SP-based resources, we only consider NetPrice (NP) models as the duty factors achieved
by LMP are too low to be competitive.
Using NP0 and NP5 models, duty factors $>$50\% can
be achieved, producing competitive and even superior cost-performance (see
Figure 
\ref{fig:scaling}).
Overall, while NP0 systems match Ctr resources, the
NP5 systems outperform, with Ctr+1Z 20\% more cost-effective than 2Ctr.
Ctr+4Z is 34\% superior to 5Ctr.

In conclusion, for our baseline (power price=\$60/MWh, and computing 
price=1x), scaling with SP-based intermittent resources 
is more cost-effective than traditional resources.  
Intermittent resources using NetPrice models are useful and beneficial to HPC. 

\begin{figure*}[tb]
\centering
\vspace{-0.1in}
\includegraphics[width=6.4in]{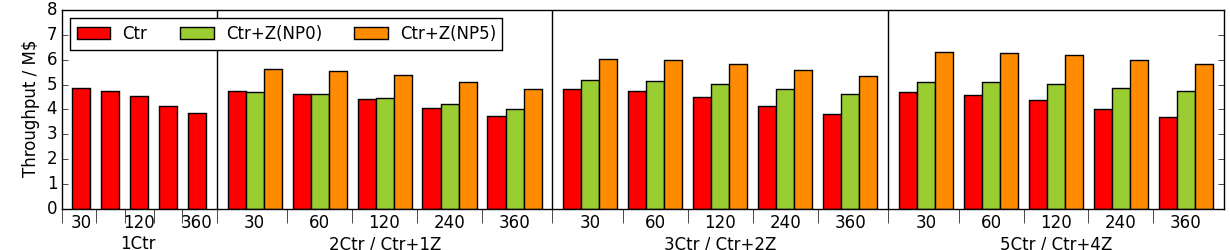}
\vspace{-0.1in}
\caption{Throughput/M\$ vs. System scale vs. Power price.  System scale increases from left to right.  Power
price from \$30/MWh to \$360/MWh.}
\vspace{-0.08in}
\label{fig:scaling_power}
\end{figure*}

\begin{figure*}[tb]
\centering
\vspace{-0.1in}
\includegraphics[width=6.4in]{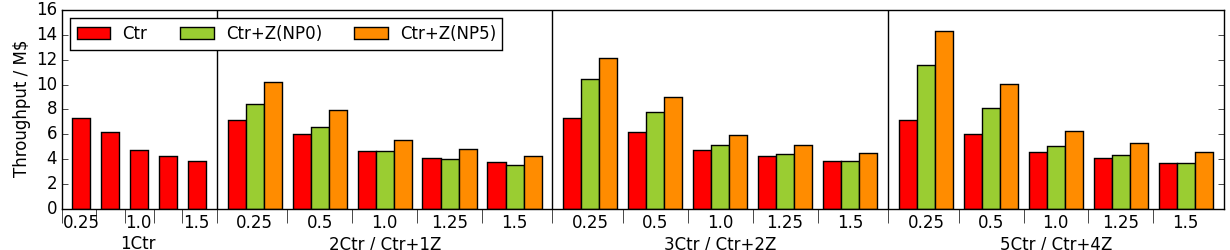}
\vspace{-0.1in}
\caption{Throughput/M\$ vs. System scale vs. Compute HW price.  System
  scale increases from left to right.  Compute HW price ranges from
  0.25x to 1.5x}
\vspace{-0.08in}
\label{fig:scaling_compute}
\end{figure*}

\begin{figure*}[tb]
\centering
\vspace{-0.1in}
\includegraphics[width=6.4in]{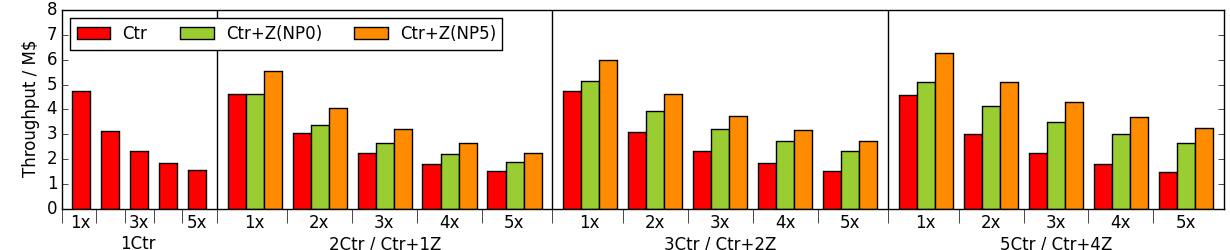}
\vspace{-0.1in}
\caption{Throughput/M\$ vs. System scale vs. System Power Density.  System scale increases from left to right.  System power density ranges from 1x to 5x}
\vspace{-0.08in}
\label{fig:scaling_density}
\end{figure*}

\subsection{The Impact of Power Price}

The price of power varies widely from country to country, and even the region
within a country \cite{power-prices}.  For example, at the high end,
the average retail price for power in Germany in 2011 was \$0.35/kwh =
\$350/MWh and in Denmark, the greenest country in Europe, \$0.41/kwh =
\$410/MWh.  On the other side of the globe, Japanese power costs
\$0.25/kwh = \$250/MWh, but the largest asian nations China and India
have much lower \$0.08/kwh = \$80/MWh prices.  The United States falls
in the middle \$0.12/kwh = \$120/MWh.  For completeness, we consider a
range down to the wholesale prices in the US, which are \$30/MWh for
the Midcontinent ISO (MISO).

Beginning at the left of Figure
\ref{fig:scaling_power}, in all cases as the power price increases, 
system throughput/M\$ decreases.  The Ctr+Z systems match the 2Ctr performance with 
NP0 at low power prices, and outperforms slightly at a power price typical 
for Japan, Germany or Denmark.  With NP5, the advantage is
more than 29\%.  
Adding more ZCCloud resources increases benefit.  With a larger fraction of 
intermittent resources, the ZCCloud advantage grows, and
at 4x ZCCloud resources, the throughput/M\$ of Ctr+4Z(NP5) is 55\% 
higher than 5Ctr.  There is a slight decline
as power price increases, but since ZCCloud resources in 
Ctr+Z exploit zero-cost stranded power, the decay is much 
slower, and advantage increases.
 

So higher power prices make NetPrice-based ZCCloud resources more
cost-effective than traditional datacenter resources.  NP0-based
resources are comparable, and long stranded power intervals produced by
NP5-based resources enable them to outperform
traditional resources.  In high price power countries, ZCCloud systems
have a major cost-performance advantage, about 55\%.  In more moderate-price 
regions the benefit can be 10-30\%, growing with larger systems.

\subsection{The Impact of Compute Hardware Price}

Next we consider how changing compute hardware costs affect ZCCloud benefits.  
At the End of Moore's Law \cite{NoMoreMoore13,BorkarChien11}, 
the cost of hardware may change significantly.  If feature
scaling slows or ends --  commoditizing processors -- one could imagine
a 4x reduction (Intel processors at ARM prices).  On the other hand,
if slow progress continues, with heroic multi-patterning lithography,
prices could creep upward, perhaps 1.5x \cite{intel-chip-price}.  So, we consider a range of
prices.

Varying hardware price produces a consistent trend across all system sizes and types;
with cost increases, throughput/M\$ decreases (see Figure
\ref{fig:scaling_compute}).  The 1.0x scenarios (Baseline) match data in prior graphs,
so we consider first decreasing hardware prices.  In such
scenarios, the advantage of ZCCloud grows dramatically, 
Ctr+1Z(NP5) achieves
42\% superior throughput/M\$ than 2Ctr at 0.25x hardware cost.
For larger systems the advantage grows, reaching 
97\% for Ctr+4Z(NP5) versus 5Ctr.  On the other hand, if hardware
cost increases, ZCCloud's benefits suffer from a lower duty factor, and
poor utilization of expensive hardware.  
At 1.5x hardware price, Ctr+Z systems are comparable with Ctr resources.  
In conclusion, Ctr+Z benefits more from falling hardware prices than Ctr.  



\subsection{The Impact of Growing System Power Density}

The growth of supercomputer power is a well-documented trend \cite{top500,exascale-challenges}
with the power for a state-of-the-art
supercomputing system at a given price (\$150M for the US DOE
CORAL systems \cite{Summit,Aurora}) is increasing by 3x per five-year generation.
Note
that each generation has a much greater computation rate (nearly 20x),
We explore the impact of this increasing power {\it density} MW/\$,
and scale up the workload by assuming each job is scaled up in 
proportion to the computation rate increase, maintaining
constant job throughput.

With increasing power density, all systems see a steady decrease in
throughput/M\$ (see Figure \ref{fig:scaling_density}).  
This trend is pronounced for all system 
scales and all configurations.
A decrease in job throughput may still be an increase
in computational throughput, as the jobs are scaling up.
However the ZCCloud systems decline less precipitously, so the 
relative difference grows, and the growth of power density makes ZCCloud more
attractive.  While throughput/M\$ of Ctr+1Z(NP0) is almost the same as that of 2Ctr for 1x 
density, its 25\% better than 2Ctr for 5x density. Ctr+1Z(NP5) is 
20\% more cost-effective than 2Ctr for 1x density, and 50\% 
better than 2Ctr for 5x density.  Scaling to more intermittent resources 
further improves cost-effectiveness.  For example, for 1x density, Ctr+4Z(NP5) achieves 34\% higher throughput/M\$ than 5Ctr; and for 5x density, the 
throughput/M\$ of Ctr+4Z(NP5) is remarkably 116\% higher than 5Ctr.  

With higher system power density, all system
configurations see declining cost-effectiveness, 
the use of stranded power with NP5
ZCCloud advantages that grow with power density and scale,
reaching as much as 116\% better.

\section{ZCCloud at Extreme Scale}
\label{sec:exascale}

\begin{figure}[bt]
\centering
\vspace{-0.1in}
\includegraphics[width=3in]{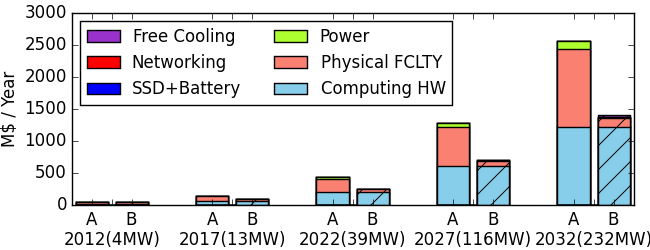}
\vspace{-0.1in}
\caption{TCO breakdown for Traditional (A) vs. ZCCloud (B).}
\vspace{-0.08in}
\label{fig:exascale_breakdown_ambitious}
\end{figure}


%
%
%
%


There are major questions how to scale to exascale and beyond, as the growth of
power and required physical infrastructure are daunting.  
Here we consider two approaches: (1) traditional datacenters (Scenario A)
and, (2) ZCCloud - adding stranded power systems to complement a 4MW base
system.



Figure \ref{fig:exascale_breakdown_ambitious} applies our TCO model to
extreme scaling, and using a constant cost per unit for IT hardware,
datacenter, and electrical power.  The results follow those in
Section \ref{sec:cost}.

To create a baseline, we document performance and power for leading DOE
systems: 2012 \cite{mira} and 2017 \cite{Aurora}, and project based
on geometric growth for 2022, 2027, and 2032 in Table
\ref{table:generations}.  The numbers are daunting, despite building
in 7x energy/op improvement every 5 years, and
conservative compared to Horst Simon's
empirical scaling model \cite{SimonScaling}.

\begin{table}[t!]
\renewcommand{\arraystretch}{1.3}
\vspace{-0.1in}
\caption{Projected Performance and Power for Top DOE Supercomputer Systems}
\label{table:generations}
\vspace{-0.1in}
\centering
{\small
\begin{tabular}{|l|r|r|r|r|r|}
\hline
Year & 2012 & 2017 & 2022 & 2027 & 2032 \\
\hline \hline
Peak PF & 10 \cite{mira} & 200 \cite{Aurora} & 4K & 80K & ? \\
\hline
MW & 4 \cite{mira} & 13 \cite{Aurora} & 39 & 116 & 232 \\
\hline \hline
\multicolumn{6}{c}{\bf Horst Simon Model \cite{SimonScaling}} \\
\hline
Peak GF/kW & 2.2K & 4.2K & 6.2K & 8.2K & 10.2K \\
\hline
MW & 2 & 48 & 645 & 9.8K & ? \\
\hline
\end{tabular}
}
\vspace{-0.08in}
\end{table}

\begin{figure}[bt]
\centering
\vspace{-0.1in}
\includegraphics[width=2.8in]{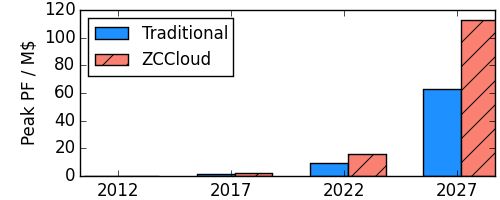}
\vspace{-0.1in}
\caption{Peak PFLOPS/M\$ for traditional vs. ZCCloud, scaling TCO.}
\vspace{-0.08in}
\label{fig:exascale_ambitious}
\end{figure}

Considering differences in TCO growth, we can compare
cost-effectiveness in achieving ``high peak performance'', peak petaflops/million-dollars TCO (see Figure
\ref{fig:exascale_ambitious}).  By shedding power cost and significant power distribution/cooling and
other physical infrastructure, the ZCCloud approach
have significant advantages for enabling capability computing.
Specifically, TCO for the 39MW, 116MW and 232MW traditional systems
are \$430M, \$1,300M, and \$2,550M/year respectively.  ZCCloud
approach mitigates both power and datacenter elements of TCO growth,
reducing TCO by 41\% at 39MW and more than 45\% at 232MW.  These
estimates overstate likely hardware costs (constant HW cost/MW), but any
reductions shift TCO balance to power and datacenter costs and thus
further increase the advantage of the ZCCloud approach.



\begin{figure}[bt]
\centering
\vspace{-0.1in}
\includegraphics[width=2.8in]{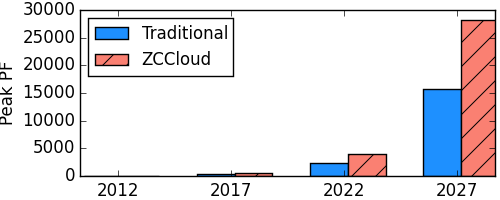}
\vspace{-0.1in}
\caption{Peak PFLOPS for traditional vs. ZCCloud, \$250M/year TCO limit.}
\vspace{-0.08in}
\label{fig:exascale_limited_ambitious}
\end{figure}

Supercomputer systems are often limited to a fixed budget, so we consider
growing TCO, and compute the maximum size system that can be built and operated
given a TCO of \$250M/year (Figure \ref{fig:exascale_limited_ambitious}).
Here the cost efficiency for peak performance of ZCCloud enables  
80\% greater peak PFLOPS.

\begin{figure}[bt]
\centering
\vspace{-0.1in}
\includegraphics[width=2.8in]{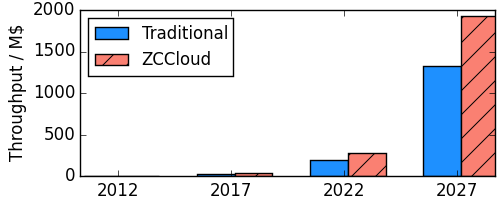}
\vspace{-0.1in}
\caption{Overall cost-competitiveness for theoretical throughput: traditional vs. ZCCloud, scaling TCO.}
\vspace{-0.08in}
\label{fig:extreme_scale_throughtput_cost_eff}
\end{figure}

Finally, we consider cost-effectiveness on a throughput basis. 
We we combine the
productivity of ZCCloud hardware at duty factors feasible on stranded
power (80\%) with the TCO scaling to compute throughput in
jobs/million-dollars of TCO (Figure
\ref{fig:extreme_scale_throughtput_cost_eff}), showing
that ZCCloud approaches can are 45\% more cost-effective.



\section{Discussion and Related Work}
\label{sec:discussion}





Numerous researchers have studied integrating renewables into
datacenters and  maximizing green power usage \cite{goiri2013parasol,
li2012iswitch,haque2015greenpar}.  More ambitious efforts even 
``follow the renewables'' \cite{berral2014building} distributing
computing resources geographically to increase 
green power usage.   ZCCloud differs from these efforts, using 
no grid power (brown or otherwise), exploiting only un-economic stranded power
at the point of generation, a dispatchable load \cite{KYZC2016,chien2015zero} that
may even benefit grid stability.

Improving data-center energy-efficiency is an area of intensive
research.
For example, under-provisioning reduces grid power supply to
the average compute power demand instead of peak
\cite{zhou2015underprovisioning}.  Dynamic power management selects
power source based on the load \cite{liu2015heb}, and energy-aware job
schedulers migrate jobs and shift peak load to achieve higher
efficiency \cite{haque2015greenpar}.  These approaches achieve higher
power efficiency at some trade in system performance and quality of
service, such as lowering CPU frequency and deferring jobs. In
contrast, ZCCloud is designed as an extension to supercomputing
systems to scale up throughput and performance.

Distributed systems researchers have long sought to exploit
intermittent computing resources, notably Peer-to-peer \cite{P2PBook},
desktop grid \cite{Entropia, SETI}, and Condor \cite{condor}, where
computing resources are generally single machines, loosely-coupled, and
volatile.  Cloud providers have also operated revokable computing
services.  Amazon's spot instances \cite{spot-instance} enable cloud
uses to bid on EC2 instances with a lower cost, and revoke resource
when its price is higher than the user's bid price. Google Compute
Engine (GCE) provides analogous capability in a preemptible VM
instance \cite{gce-preemptible}.  A number of research studies explore
how to make these volatile resources more useful for online services
requiring continuous availability \cite{hpdc15-spot1, hpdc15-spot2,cycle-computing}.  We believe these efforts may have the potential
to create high-value cloud services based on 
volatile ZCCloud resources.

Researchers have proposed migrating workloads around the world to
reduce electricity cost at Akamai \cite{Qureshi09}, and
``follow-the-sun'' techniques to increase the fraction of renewable
energy use.  Later techniques combine optimization for renewable use
while meeting response time requirements \cite{berral2014building}.
All of these techniques suffer from TCO's dominated by the 
capital costs for compute hardware and physical infrastructure, and
thus migratory techniques that produce low hardware resource utilization
are generally not cost-effective.  Our work on ZCCloud assesses achievable
duty factors and overall costs, showing the ZCCloud can be more cost-effective
than a single traditional supercomputing center facility.

Our work builds on cost-model research for traditional and
green datacenters.  Researchers have proposed frameworks that model
and optimize the cost/performance of a network of datacenters,
considering capex and opex, and regional variations in
electricity, networking, and real estate \cite{le2011intelligent,wang15grid}.

\section{Summary and Future Work}
\label{sec:conclusion}
\label{sec:summary}
We explore cost-effectiveness of a novel
approach to scaling supercomputing.  ZCCloud uses
stranded power, and locates computing hardware at selected 
renewable generation sites to reduce physical building and power distribution
costs.
MISO energy market statistics show that
sufficient stranded power is available
support large-scale
computing resources with duty factors as high as 80\%.
Simulations
of production HPC workload show that
intermittent resources can effectively increase system capabilities.

Our study of ZCCloud cost-effectiveness shows that
stranded-power based computing resources 
can save up to 45\% of capital costs at a given scale
and improve cost-effectiveness by more than 100\%.  
Study of future extreme scale systems 
reveal significant advantages due to reduced
power and physical infrastructure, ZCCloud extension can reduce 
capital costs significantly.

Beyond simulation, one might build a ZCCloud prototype
to prove out the costs and effectiveness on real workloads.
Our first 108-core prototype is driven by historical MISO stranded power data,
and is executing Open Science Grid 
(OSG) \cite{osg} workloads.  Our second planned prototype will be deployed in 
the next three months, and includes 78 nodes (2,500 cores), and we will
explore OSG workloads and batch workloads from a local supercomputing 
center.  Beyond that, we will build a physical container prototype,
and deploy it at a carefully selected wind farm with ample stranded
power.
Another
interesting direction is to explore support for computing services
beyond batch parallel computing.




%

\appendix
\section{TCO Model Details}
\label{sec:tco-details}


The amortized costs in Table \ref{table:tco_parameters} are calculated 
using the following formulas:
\begin{equation}
C_{component}=\frac{r\cdot CapEx_{component}}{1-(1+r)^{-l}}
\end{equation}
\begin{equation}
CapEx_{component}=Price_{component}\cdot Size_{component}
\end{equation}
Where $CapEx_{component}$ is the capital expense of the component, 
$r$ is the cost of capital, 3\%, and 
$l$ is the amortization period (in years).  
Table \ref{table:detail_cost} includes full detail.

\begin{table}[hbt]
\renewcommand{\arraystretch}{1.3}
\vspace{-0.1in}
\caption{Detailed Parameters of the Cost Model}
\label{table:detail_cost}
\vspace{-0.1in}
\centering
{\small
\begin{tabular}{l|l|l|l}
\hline
Component & Price & Size & Amortization\\
\hline
\hline
Compute & \$24M/MW & 4MW & 5 years\\
\hline
Network & \$13k/mile & 500 miles & 10 years \\
\hline
SSD & \$0.67/GB & 2PB & 5 years\\	
\hline
Battery & \$350/kWh & 1MWh & 5 years\\
\hline
Container & \$5M/MW & 4MW & 12 years\\
\hline
Free Cooling & \$700k/MW & 4MW & 10 years\\
\hline
\end{tabular}
}
\vspace{-0.1in}
\end{table}

We present all elements from \cite{le2011intelligent} 
in Table \ref{table:bianchini_cost}; our cost model 
omit those with contributions generally less that $<$1\% 
including water, connection and land. 

\begin{table}[hbt]
\renewcommand{\arraystretch}{1.3}
\vspace{-0.1in}
\caption{Full Cost Model from \cite{le2011intelligent}}
\label{table:bianchini_cost}
\vspace{-0.1in}
\centering
{\small
\begin{tabular}{l|p{4cm}|p{2.2cm}}
\hline
Elements & Description & Contribution\\
\hline \hline
\multicolumn{3}{c}{\bf Components included in our simple model} \\
\hline
Servers & Computing hardware costs & 44.9\% - 56.9\% \\
\hline
Building & Machine room, cooling and power infrastructures & 23.6\% - 34.5\% \\
\hline
Power & Electricity costs & 9.2\% - 17.8\% \\
\hline
Networking & Internal networking costs &  0.7\% - 1.1\% \\
\hline \hline
\multicolumn{3}{c}{\bf Components omitted in the simple model} \\
\hline
Connection & Connection to power grid and Internet backbone & 0.1\% - 4.7\% \\ 
\hline
Water & Water used for cooling & 0.1\% - 1.2\% \\
\hline
Land & Cost of real-estate & 0.7\% - 2.4\% \\
\hline
\end{tabular}
}
\vspace{-0.1in}
\end{table}



\def\UrlBreaks{\do\/\do-\do\.}
\bibliographystyle{IEEEtran}
\bibliography{references}

\end{document}